\documentclass[epj,twocolumn]{webofc}
\usepackage[varg]{txfonts}   

\usepackage{cuted}
\usepackage{geometry}
\usepackage{url}
\woctitle{International Workshop on Radiopure Scintillators RPSCINT 2013}
\begin{document}
\title{Search for 2$\beta$ decay of $^{116}$Cd with the help of enriched $^{116}$CdWO$_4$ crystal scintillators}
%
%

\author{D.V.~Poda\inst{1}\fnsep\thanks{\email{poda@kinr.kiev.ua}} \and
        A.S.~Barabash\inst{2} \and
        P.~Belli\inst{3} \and
        R.~Bernabei\inst{3,4} \and
        F.~Cappella\inst{5,6} \and
        V.~Caracciolo\inst{7} \and
        S.~Castellano\inst{7} \and        
        D.M.~Chernyak\inst{1} \and        
        R.~Cerulli\inst{7} \and        
        F.A.~Danevich\inst{1} \and        
        S.~d'Angelo\inst{3,4} \and        
        A.~Incicchitti\inst{5,6} \and        
        V.V.~Kobychev\inst{1} \and        
        S.I.~Konovalov\inst{2} \and        
        M.~Laubenstein\inst{7} \and        
        R.B.~Podviyanuk\inst{1} \and        
        O.G.~Polischuk\inst{1,6} \and        
        V.N.~Shlegel\inst{8} \and        
        V.I.~Tretyak\inst{1} \and        
        V.I.~Umatov\inst{2} \and        
        Ya.V.~Vasiliev\inst{8}
        }

\institute{Institute for Nuclear Research, MSP 03680 Kyiv, Ukraine 
\and
           Institute of Theoretical and Experimental Physics, 117259 Moscow, Russia 
\and
           Dipartimento di Fisica, Universit\`a di Roma ``Tor Vergata'', I-00133 Rome, Italy
\and
           INFN, sezione di Roma "Tor Vergata", I-00133 Rome, Italy
\and
           Dipartimento di Fisica, Universit\`a di Roma ``La Sapienza'', I-00185 Rome, Italy
\and
           INFN, sezione di Roma "La Sapienza", I-00185 Rome, Italy
\and
           Laboratori Nazionali del Gran Sasso, INFN, I-67100 Assergi (AQ), Italy
\and
           Nikolaev Institute of Inorganic Chemistry, 630090 Novosibirsk, Russia
          }

\abstract{%
  Cadmium tungstate crystal scintillators enriched in $^{116}$Cd to 82\% ($^{116}$CdWO$_4$, total mass of $\approx$1.2 kg) 
  are used to search for 2$\beta$ decay of $^{116}$Cd deep underground at the Gran Sasso National Laboratory of the 
  INFN (Italy). The radioactive contamination of the $^{116}$CdWO$_4$ crystals has been studied carefully to 
  reconstruct the background of the detector. The measured half-life of $^{116}$Cd relatively to 2$\nu$2$\beta$  
  decay is $T^{2\nu2\beta}_{1/2}$ = [2.8 $\pm$ 0.05(stat.) $\pm$ 0.4(syst.)] $\times$ 10$^{19}$ yr, in agreement 
	with the results 	of previous experiments.   The obtained limit on the 0$\nu$2$\beta$ decay of $^{116}$Cd 
	(considering the data of 	the last 8696 h run with an advanced background 0.12(2) counts/yr/kg/keV in the 
	energy interval $2.7-2.9$ MeV) 	is $T_{1/2} \ge 1.0 \times 10^{23}$ yr at 90\% C.L. The sensitivity of the 
	experiment to the $0\nu2\beta$ process 	is $\lim T_{1/2} = 3 \times 10^{23}$ yr at 90\% C.L. over 5 years 
	of the measurements and it can be advanced 	(by further reduction of the background by a factor $3-30$) to the 
	level of 	$\lim T_{1/2} = (0.5-1.5) \times 10^{24}$ yr for the same period of the data taking.
}
\maketitle
\section{Introduction}
\label{intro}

The rarest nuclear decay among observed --- double beta decay ($2\beta$) --- is a powerful tool to 
check physics beyond the Standard Model (SM) of particle physics.  Allowed in the SM two neutrino 
double beta decay ($2\nu2\beta$) has been registered for only 11 nuclides with a typical half-lives 
in the range of $10^{18}-10^{24}$ yr (see e.g. in \cite{DBTabl,Bar10a}). Neutrinoless double beta 
decay ($0\nu2\beta$) --- forbidden in the SM process --- should exist, if neutrino is a Majorana particle, 
thanks to recently observed non-zero masses of the neutrino \cite{PDG}. The observation of the 
$0\nu2\beta$ decay will testify the lepton number non-conservation, the Majorana nature of the neutrino, 
an absolute value and hierarchy of the neutrino masses. The $0\nu2\beta$ decay is also sensitive to 
an existence of the right-handed currents, hypothetical particles (majorons) and other effects beyond 
the SM (see the recent reviews \cite{Cre13,Ver12,Ell12,Giu12,Gom12,Rod11} and references therein).

One of the most promising candidates for the $2\beta$ decay search is $^{116}$Cd thanks to 
high $2\beta$ energy release ($Q_{2\beta}$ = 2813.50(13) keV \cite{Rah11}), 
large natural isotopic abundance ($\delta$ = 7.49\% \cite{Ber11}), 
availability of enrichment by relatively cheap centrifugal method \cite{Art97}, 
and promising theoretical estimations (see e.g. \cite{Ver12}). 
Moreover, existence of excellent detector, cadmium tungstate (CdWO$_4$) scintillator, 
already used for $2\beta$ decay searches \cite{Dan95,Dan00,Dan03} allows to realize 
``source = detector'' experiment with a high detection efficiency.  
In addition, detector based on cadmium tungstate scintillator has several features 
(see e.g. \cite{Dan03,Bar06}) which are very important for such kind of studies: 
low level of intrinsic radioactivity, good scintillation properties, 
particle identification ability, relatively low cost of the scintillation material, 
stability for long term operation.

Recently an enriched $^{116}$CdWO$_4$ crystal scintillator with a large mass (1.87 kg) 
has been developed from deeply purified raw materials, and a new high sensitive 
$2\beta$ experiment has been started in February 2011 at the Gran Sasso
National Laboratories (LNGS) of the INFN (Italy) \cite{Bar11}. The results 
of the last run performed from October 2012 till October 2013 will be reported here.

\section{Low-background experiment}
\label{sec-1}

The experiment has been performed deep underground ($\approx$ 3600 m w.e.) at the LNGS (Italy). 
The description of the detector and low background set-up at the previous stages of the experiment 
one can find in Refs. \cite{Bar11, Bar13a}. Here we outline the main features of the last stage.

Two $^{116}$CdWO$_4$ crystal scintillators (586 g and 589 g) were fixed inside 
polystyrene light-guides ($\oslash70\times194$ mm). The cavities inside the light-guides were filled with  
the Ultima Gold AB (PerkinElmer) liquid scintillator (LS). Two high purity quartz light-guides ($\oslash70\times200$ mm) 
were glued to the polystyrene light-guide on both sides. The detectors were viewed from the opposite sides 
by two low radioactive 3'' photomultipliers (PMTs, Hamamatsu R6233MOD). The detector modules, 
covered by the 3M foil to improve the light collection, were installed inside a low radioactive 
air-tight Cu box at the DAMA/R\&D set-up. Copper bricks 5 cm thick were used 
as an additional passive shield. The Cu box was continuously flushed with high purity nitrogen gas to 
remove the residual environmental radon. Outer passive shield consists of high purity Cu (10 cm thick), 
15 cm of low radioactive lead, 1.5 mm of cadmium and 4 to 10 cm of polyethylene/paraffin. The whole shield is contained 
inside a Plexiglas box, also continuously flushed by high purity nitrogen. 

An event-by-event data acquisition system based on a 1 GS/s 8 bit transient digitizer (Acqiris DC270) 
was used to record time of each event and scintillation pulse-profiles. The energy scale and the resolution 
of the detector was calibrated by using the reference $^{22}$Na, $^{60}$Co, $^{137}$Cs, and $^{228}$Th 
$\gamma$ sources. The energy resolution of the detector was FWHM 4.3(4)\% at $Q_{2\beta}$ of $^{116}$Cd. 

\section{Results and discussion}
\label{sec-2}

\begin{figure}
\centering
\includegraphics[width=9cm]{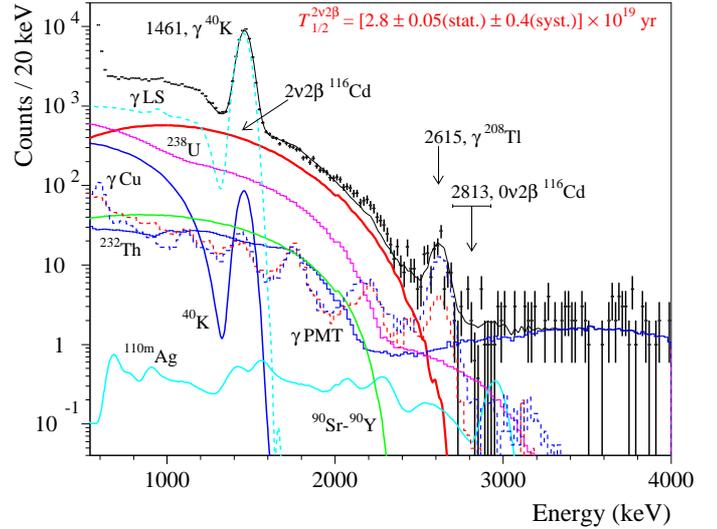}
\caption{(Color online) The energy spectrum of  $\gamma(\beta)$ events measured with 
the $^{116}$CdWO$_4$ scintillation detector over 8696 h in the low-background set-up 
(points with error bars) together with the fit (solid black line). The main components 
of the background are also shown: $2\nu2\beta$  decay of $^{116}$Cd, the distributions 
of internal $^{40}$K,  $^{90}$Sr-$^{90}$Y, $^{232}$Th, $^{238}$U, and cosmogenic $^{110m}$Ag, 
the contribution from external $\gamma$ quanta from the details of the detector (``LS'', ``PMT'') 
and set-up (``Cu'') under these experimental conditions. The energies are in keV.}
\label{fig-1}    
\end{figure}

The data at this stage were collected over 8696 h. The applied data analysis is similar to the one 
described in \cite{Bar11}. The energy spectrum of $\gamma(\beta)$ events accumulated 
with the $^{116}$CdWO$_4$ detector and selected off-line by the pulse-shape and the front edge analyzes 
is shown in Fig.~\ref{fig-1}. The background model was built from possible components 
like internal $^{40}$K, $^{90}$Sr-$^{90}$Y, $^{110m}$Ag, radionuclides from 
U/Th chains, external $\gamma$ from the details of the set-up, $2\nu2\beta$ decay of $^{116}$Cd 
 simulated by using the EGS4 \cite{EGS4} and Decay0 \cite{Decay0} codes. 
The radiopurity of the $^{116}$CdWO$_4$ crystals is reported in \cite{Bar11,Pod13,Dan13}. 
The radioactive contamination of the PMTs \cite{Ber12} and the sample of Ultima Gold AB 
liquid scintillator has been measured at the STELLA HPGe facility of the LNGS (Italy). 
The radiopurity of Cu has been derived from the fit of the background spectrum and it is comparable 
with the results of the low background measurements with the Cu samples (see e.g. \cite{Heu06}).
Table~\ref{tab-1} contains radiopurity level of the main sources of the background. 

\begin{table*}
\centering
\caption{Radioactive contamination of the details of the $^{116}$CdWO$_4$ 
 detector ($^{116}$CdWO$_4$ crystals No.1 and No.2, liquid scintillator LS, and PMTs) 
 and the inner part of the low-background DAMA/R\&D set-up made from Copper.}
\begin{tabular}{llllll}
 \hline
 Nuclide    & \multicolumn{5}{c}{Activity (mBq/kg)} \\
\cline{2-6}
 ~ & $^{116}$CdWO$_4$ No.1 & $^{116}$CdWO$_4$ No.2 & LS & PMT      & Copper \\
 \hline
$^{228}$Ra  & ~           & ~           & ~             & 120(20)  & ~                    \\
$^{228}$Th  & 0.031(3)    & 0.054(5)    & ~             & 83(17)   & 0.025(14)          \\
\hline
$^{238}$U   & 0.5(2)      & 0.7(2)      & ~             & ~        & ~              \\
$^{226}$Ra  & $\leq0.005$ & $\leq0.005$ & ~             & 430(60)  & 0.011(3) \\
$^{210}$Po  & 0.6(2)      & 0.8(2)      & ~             & ~        & ~                   \\
\hline
$^{40}$K    & $\leq0.9$   & $\leq0.9$   & 1600(200)     & 540(160) & 0.04(1)   \\
$^{90}$Sr-$^{90}$Y & $\leq0.1$   & $\leq0.1$   & ~             & ~        & ~                  \\
$^{110m}$Ag & $\leq0.7$   & $\leq0.7$   & ~             & ~        & ~                  \\
 \hline
\end{tabular}
 \label{tab-1}
 \end{table*}

Fit of the background spectrum and the main components of the background are also 
presented in Fig.\ref{fig-1}. The fit gives the number of the $2\nu2\beta$ events of 
$^{116}$Cd as $S_{2\nu2\beta} = (34927 \pm 614$) counts for the total distribution. 
The signal to background ratio in the energy range of $1.6-2.4$ MeV is $\approx$ 2:1. 
Taking into account the efficiency of the applied 
pulse-shape discrimination procedure ($\eta_{PSD}$ = 0.92), 
selection of the $^{212}$Bi-Po events ($\eta_{BiPo}$ = 0.99), 
detection efficiency to the process searched for ($\eta_{2\nu2\beta}$ = 0.986),
the number of $^{116}$Cd nuclei in the used enriched $^{116}$CdWO$_4$ crystals 
(1162.25 g, enrichment 82.2\%, $N_{116} = 1.584 \times 10^{24}$ nuclei), 
time of measurements ($t$ = 0.992 yr), one can calculate the half-life of $^{116}$Cd 
relatively to the $2\nu2\beta$ decay by using the following formula: 
$T^{2\nu2\beta}_{1/2} = \ln2 \cdot \eta_{PSD} \cdot \eta_{BiPo} \cdot \eta_{2\nu2\beta}  
\cdot N_{116} \cdot t$ / $S_{2\nu2\beta}$.  The derived half-life 
$T^{2\nu2\beta}_{1/2}$ = [2.8 $\pm$ 0.05(stat.) $\pm$ 0.4(syst.)] $\times$ 10$^{19}$ yr 
is in agreement with the results of the previous experiments (see Table~\ref{tab-2}) 
and the recent world average value $(2.8 \pm 0.2) \times 10^{19}$ yr \cite{Bar10b}. 
	
\begin{table*}
\centering
\caption{The half-lives of $2\nu2\beta$ decay of $^{116}$Cd (g.s.$\rightarrow$g.s.) 
measured in the $2\beta$ experiments. The signal to background ratio (S:Bg), main 
experimental features, and the total exposure are also presented.}
\begin{tabular}{lllcl}
\hline
$T^{2\nu2\beta}_{1/2}$ ($10^{19}$ yr)         & S:Bg & Experiment & Exposure (kg$\times$yr) & Year [Ref.]\\
\hline
=2.6 $^{+0.9}_{-0.5}$                          & 1:3  & ELEGANT V, $^{116}$Cd foils          & 0.02 & 1995 \cite{Eji95} \\
=2.7 $^{+0.5}_{-0.4}$(stat.) $^{+0.9}_{-0.6}$(syst.) & 1:1 & Solotvina, $^{116}$CdWO$_4$ scintillators & 0.04 & 1995 \cite{Dan95}\\
=3.75 $\pm$ 0.35(stat.) $\pm$ 0.21(syst.)     & 4:1  & NEMO-2, $^{116}$Cd foils          & 0.11 & 1995 \cite{Arn95,Arn96} \\
=2.6 $\pm$ 0.1(stat.) $^{+0.7}_{-0.4}$(syst.)  & 4:1  & Solotvina, $^{116}$CdWO$_4$ scintillators & 0.18 & 2000 \cite{Dan00} \\
=2.9 $\pm$ 0.06(stat.) $^{+0.4}_{-0.3}$(syst.) & 3:1  & Solotvina, $^{116}$CdWO$_4$ scintillators & 0.48 & 2003 \cite{Dan03} \\
=2.88 $\pm$ 0.04(stat.) $\pm$ 0.16(syst.)     & 10:1 & NEMO-3, $^{116}$Cd foils        & 1.95 & 2010 \cite{Pah10,Tre11,Pah12} \\
=2.5 $\pm$ 0.5                                & 2:1  & Present, $^{116}$CdWO$_4$ scintillators & 1.04 & 2012 \cite{Bar13a} \\
=2.8 $\pm$ 0.05(stat.) $\pm$ 0.4(syst.)       & 2:1  & Present, $^{116}$CdWO$_4$ scintillators & 1.15  & 2013  \\
\hline
\end{tabular}
\label{tab-2}       
\end{table*}


In case of the $0\nu2\beta$ decay of $^{116}$Cd to the ground state of $^{116}$Sn we should see in the 
experimental spectrum a peak at the energy $\approx$ 2.8 MeV. There is no such peculiarity in the 
spectrum, therefore only lower half-life limit on the process can be set according to formula: 
$\lim T^{0\nu2\beta}_{1/2} = \ln2 \cdot \eta_{PSD} \cdot \eta_{BiPo} \cdot \eta_{0\nu2\beta} \cdot 
N_{116} \cdot t / \lim S$. Here $\eta_{0\nu2\beta}$ is the detection efficiency for the 
$0\nu2\beta$ process (equal to 0.947), $\lim S$ is the number of $0\nu2\beta$ events which can be 
excluded at a given confidence level (C.L.), and other parameters are described above.  
The fit gives the area of the $0\nu2\beta$ peak as (2.3 $\pm$ 4.4) counts. It corresponds to 
$\lim S$ = 9.2 counts at 90\% C.L. according to the Feldman-Cousins procedure \cite{Fel98}. 
Substituting all parameters, one can obtain the limit 
$T^{0\nu2\beta}_{1/2} \ge 1.0 \times 10^{23}$ yr at 90\% C.L., which is on the level of 
the Solotvina experiment \cite{Dan03} (see Table~\ref{tab-3}). 

\begin{table*}
\centering
\caption{The half-life limits on the $0\nu2\beta$ decay of $^{116}$Cd (g.s.$\rightarrow$g.s.) 
established in the $2\beta$ experiments. The background rate in the region of interest (Bg in ROI), 
main experimental features, enrichment, and the total exposure are also presented.}
\begin{tabular}{cclccl}
\hline
$T^{0\nu2\beta}_{1/2}$ (yr) & Bg in ROI & Experiment  & Enrichment & Exposure & Year [Ref.]\\
 at 90\% C.L.           & (counts/yr/kg/keV) & ~             & (\%)       & (kg$\times$yr) & ~ \\
\hline
$\ge2.9 \times 10^{21}$ & ~    & ELEGANT V, $^{116}$Cd foils       & 90.7 & 0.02         & 1995 \cite{Eji95} \\
$\ge2.9 \times 10^{22}$ & 0.53 & Solotvina, $^{116}$CdWO$_4$ scintillators & 83.0 & 0.08 & 1995 \cite{Dan95} \\
$\ge5.0 \times 10^{21}$ & 0.02 & NEMO-2, $^{116}$Cd foils          & 93.2 & 0.11         & 1995 \cite{Arn95,Arn96} \\
$\ge7.0 \times 10^{22}$ & 0.03 & Solotvina, $^{116}$CdWO$_4$ scintillators & 83.0 & 0.18 & 2000 \cite{Dan00} \\
$\ge1.7 \times 10^{23}$ & 0.04 & Solotvina, $^{116}$CdWO$_4$ scintillators & 83.0 & 0.53 & 2003 \cite{Dan03} \\
$\ge9.4 \times 10^{19}$ & $\approx$9 & COBRA, CdZnTe semiconductors & nat. & 0.05        & 2009  \cite{Daw09} \\
$\ge1.3 \times 10^{23}$ & 0.005 & NEMO-3, $^{116}$Cd foils         & 93.3 & 1.95         & 2010 \cite{Pah10,Pah12} \\
$\ge9.2 \times 10^{20}$ & $\approx$1 & COBRA, CdZnTe semiconductors & nat. & 0.23        & 2013  \cite{Geh13} \\
$\ge1.0 \times 10^{23}$ & 0.12 & Present, $^{116}$CdWO$_4$ scintillators & 82.2 & 1.15   & 2013 \\
\hline
\end{tabular}
\label{tab-3}       
\end{table*}

One can derive the similar result for the $0\nu2\beta$ decay of $^{116}$Cd by using also the 
``1$\sigma$ approach''. For example, there are $N$ = 28 counts in the region of interest (ROI) $2.7-2.9$ MeV, 
where the detection efficiency for the $0\nu2\beta$ process is 0.849. The value $\lim S$ 
in such a case can be taken as $1.64 \cdot \sqrt{N}$ = 8.7 counts at 90\% C.L. So, 
the calculated half-life limit $T^{0\nu2\beta}_{1/2} \ge 0.97 \times 10^{23}$ at 90\% C.L. is very close 
to the one obtained from the fit of the background spectrum.

The background 0.12(2) counts/yr/kg/keV in the ROI allows us to reach a sensitivity to the half-life 
of $0\nu2\beta$ decay of $^{116}$Cd equal to $\lim T_{1/2} = 3 \times 10^{23}$ yr at 90\% C.L. over 
5 years of measurements. In a case of further reduction of the background by a factor $3-30$, 
the sensitivity can be advanced to the level of $\lim T_{1/2} = (0.5-1.5) \times 10^{24}$ yr 
(it corresponds to the effective neutrino mass  $\left\langle m_{\nu} \right\rangle = 0.4-1.4$ eV).

The main components of the background in the ROI remain 2615 keV $\gamma$ quanta of $^{208}$Tl from 
contamination of the set-up by $^{232}$Th, and $^{208}$Tl from internal contamination of the $^{116}$CdWO$_4$ 
crystals by $^{228}$Th. The background could be suppressed by removing of one PMT from each detector 
module, by installation of an additional light-guide (quartz) between the crystal and the PMT, by replacing 
the PMT by another one with a higher level of radiopurity  (e.g. Hamamatsu R11065SEL) and by recrystallization 
of the $^{116}$CdWO$_4$ crystals to decrease thorium thanks to observed low segregation of Th and Ra by cadmium 
tungstate \cite{Pod13}. The first two modifications have already been applied in October, 2013. In addition, 
the Ultima Gold LS polluted by $^{40}$K has been replaced by the Borexino LS and the new measurements are in progress. 

\section{Conclusions}
\label{sec-3}

A low-background experiment using enriched $^{116}$CdWO$_4$ scintillators (total mass of 1.16 kg, enrichment 
by $^{116}$Cd to 82\%) is in progress at the LNGS of the INFN (Italy). The exposure of the 
experiment at the last stage of the data taking (October, 2012 -- October, 2013) is 1.15 kg$\times$yr. 
The energy resolution of the $^{116}$CdWO$_4$ detector is FWHM = 4.3\% at $Q_{2\beta}$ of $^{116}$Cd, 
the background counting rate was reduced to 0.12(2) counts/yr/kg/keV in the region of interest 
$2.7-2.9$ MeV, where the $0\nu2\beta$ peak is expected. The main components of the background in this 
region remain $\gamma$ quanta of $^{208}$Tl from the contamination of the set-up 
and internal contamination of the $^{116}$CdWO$_4$ crystals by $^{228}$Th.

The half-life relatively to the $2\nu2\beta$ decay of $^{116}$Cd to the ground level of $^{116}$Sn was measured as 
$T^{2\nu2\beta}_{1/2}$ = [2.8 $\pm$ 0.05(stat.) $\pm$ 0.4(syst.)] $\times$ 10$^{19}$ yr, in 
agreement with the results of the previous experiments. The half-life limit on $0\nu2\beta$ decay of $^{116}$Cd 
to the ground state of $^{116}$Sn was established as $T^{0\nu2\beta}_{1/2} \ge 1.0 \times 10^{23}$ at 90\% C.L. 
The current level of the background restricts the sensitivity of the experiment to $0\nu2\beta$ decay of $^{116}$Cd   
to $\lim T_{1/2} = 3 \times 10^{23}$ yr at 90\% C.L. over 5 years of the measurements. 
The sensitivity can be advanced to the level of $\lim T_{1/2} \approx (0.5-1.5) \times 10^{24}$  yr (which corresponds to the 
effective neutrino mass  $m_{\nu} = 0.4-1.4$ eV) after improvement of the background by factor $3-30$. 
An R\&D to improve the background conditions of the experiment is in progress.

\section{Acknowledgments}
\label{thanks}

D.V. Poda has been partly supported by a Grant of the President of Ukraine for young scientists in 2012 
(project GP-F36-110, reg. no. 0112U008078).

%
%
%

\end{document}